\documentclass[a4paper,11pt]{article}
\usepackage{jinstpub} 
\usepackage{subfigure,dcolumn}
\usepackage{epstopdf}
\usepackage{ulem}

\usepackage{siunitx}

\usepackage{subfigure,dcolumn}
\usepackage{epstopdf}
\usepackage{blindtext}
\usepackage{siunitx}
\usepackage{graphicx}
\usepackage{subfigure}
\usepackage{bm}
\usepackage{lineno}
\usepackage{amsmath}
\usepackage{overpic}

\newcommand{\kl}{K_{L}}

\newcommand{\bcntr}{\begin{center}}
\newcommand{\ecntr}{\end{center}}
\newcommand{\beq}{\begin{equation}}
\newcommand{\eeq}{\end{equation}}
\newcommand{\beqar}{\begin{eqnarray}}
\newcommand{\eeqar}{\end{eqnarray}}
\newcommand{\bitm}{\begin{itemize}}
\newcommand{\benu}{\begin{enumerate}}

\newcommand{\bitmb}{\begin{itemize}}
\newcommand{\benub}{\begin{enumerate}}
\newcommand{\eitm}{\end{itemize}}

\newcommand{\bfrm}{\begin{frame}}
\newcommand{\efrm}{\end{frame}}
\newcommand{\bct}{\begin{center}}
\newcommand{\ect}{\end{center}}
\newcommand{\bclm}{\begin{columns}}
\newcommand{\eclm}{\end{columns}}
\newcommand{\bpic}{\begin{overpic}}
\newcommand{\epic}{\end{overpic}}
\newcommand{\bblk}{\begin{block}}
\newcommand{\eblk}{\end{block}}

\newcommand{\eenu}{\end{enumerate}}

\newcommand{\ps}{\si{\pico\second}}

\newcommand{\nm}{\si{\nano\metre}}
\newcommand{\um}{\si{\micro\metre}}
\newcommand{\mm}{\si{\milli\metre}}
\newcommand{\cm}{\si{\centi\metre}}

\newcommand{\V}{\si{\volt}}
\newcommand{\mV}{\si{\milli\volt}}

\newcommand{\A}{\si{\ampere}}

\newcommand{\pF}{\si{\pico\farad}}

\newcommand{\ns}{\si{\nano\second}}

\newcommand{\ohm}{\si{\ohm}}

\newcommand{\W}{\si{\watt}}

\newcommand{\rad}{\si{\radian}}

\newcommand{\khz}{\si{k\Hz}}
\newcommand{\mhz}{\si{\mega\Hz}}
\newcommand{\ghz}{\si{\giga\Hz}}

\newcommand{\Trise}{\tau_{\rm rise}}

\newcommand{\gev}{\hbox{GeV}}

\def\Journal#1#2#3#4{{#1} {\bf #2} (#4) #3}

\def\NIMA{Nucl. Instrum. Methods A}

\def\NST{Nucl. Sci. Tech.}

\newcommand{\Rmnum}[1]{\expandafter\@slowromancap\romannumeral #1@}
\makeatother

\title{Development of a time calibration system for the KLM upgrade in the Belle II Experiment}

\author[a]{Ziyu Liu, }
\author[b]{Xiyang Wang, }
\author[b]{Shiming Zou, }
\author[b,1]{Xiaolong Wang\note{Corresponding author}, }
\author[a]{Junhao Yin, }
\author[a]{Minggang Zhao}
\affiliation[a]{School of Physics, Nankai University, 
94 Weijin Road, Tianjin, China}
\affiliation[b]{Institute of Modern Physics, Fudan University, 
220 Handan Road, Shanghai, China}

\emailAdd{wangxiaolong@ihep.ac.cn}

\abstract{
To meet the stringent time calibration requirements for the Belle II experiment upgrade, particularly for its large-size $\kl$ and Muon Detector comprising tens of thousands of scintillator channels with time resolutions better than $100~\ps$, we developed a compact and high-speed time calibration system. The system utilizes a laser diode as light source, integrated with a fast pulse laser drive circuit that employs high-speed switching GaN FETs and gate drivers. A prototype was constructed and rigorously evaluated using scintillators, achieving timing resolutions of about $13~\ps$ for a single calibration channel. Furthermore, internal deviations among calibration channels were analyzed, with most measurements remaining within $250~\ps$. These results highlight the system’s precision, scalability, and suitability for large-scale particle physics experiments.}


\begin{document}
\maketitle
\flushbottom

\section{Introduction}
\label{sec:intro}

The Belle II detector~\cite{B2TDR} is a large spectrometer operating at the SuperKEKB accelerator~\cite{SKB}, a high-luminosity electron-positron collider with a center-of-mass energy of $10.58~\gev$. Its outermost subsystem, the $\kl$ and Muon (KLM) Detector, is designed to identify muons and $\kl$ mesons. To address the increasing luminosity of SuperKEKB, an upgrade to the KLM detector is under consideration, with one proposed solution being the use of scintillators featuring long attenuation lengths for improved time resolution~\cite{B2CDR}. This enhancement aims to strengthen the detector’s background veto capabilities and enable the determination of the momentum of a $\kl$ meson based on accurate Time-of-Flight (TOF) measurement~\cite{HTR}. Variations among the scintillator channels of KLM, arising from differences in electronic components, cables, or geometric configurations, can significantly impact the accuracy of the timing measurements. To ensure precise timing evaluations with an upgraded KLM, a dedicated calibration system is essential.

TOF detectors utilizing scintillators typically comprise hundreds or thousands of detection channels. Laser calibration systems have been widely adopted in TOF detectors based on scintillators. For instance, the TOF detector in the BESIII experiment employs a PicoQuant PDL 800-B pulsed laser driver coupled with an LDH-P-C-440M laser diode head~\cite{bestof}, achieving timing resolutions of $67~\ps$ in the endcap and $87~\ps$ in the barrel with a configuration of two scintillator layers. Similarly, the pTC in MEG II~\cite{meg2} and the BAND in CLAS12~\cite{laser2020} utilize laser-based methods, achieving resolutions of $41~\ps$ and $89~\ps$, respectively, by introducing light into the center of the scintillators with an optical fiber.


To achieve a momentum resolution of $\sim 10\%$ for $\kl$ meson, the proposed upgraded KLM detector must attain a time resolution better than $100~\ps$~\cite{B2CDR,HTR}. It will be equipped with tens of thousands of compact  channels of highly transparent, elongated scintillator strips and silicon photomultiplier (SiPM) arrays. Our research and development efforts have demonstrated a time resolution of $70 \pm 7~\ps$ with an $135~\cm$-long scintillator strip, or $47\pm 2~\ps$ with a $50~\cm$-long strip~\cite{HTR}. Given the stringent requirements for time calibration in such a large-scale KLM detector, the development of a compact and flexible time calibration system is essential.

In this study, we present a novel time calibration system that utilizes a laser diode as the light source, integrated with a high-speed pulse laser drive circuit that employs a Gallium nitride field effect transistors (GaN FET) and a gate driver.

\section{Concept of time calibration}
\label{section2}

\begin{figure}[htbp]
\centering
\includegraphics[width=0.6\textwidth]{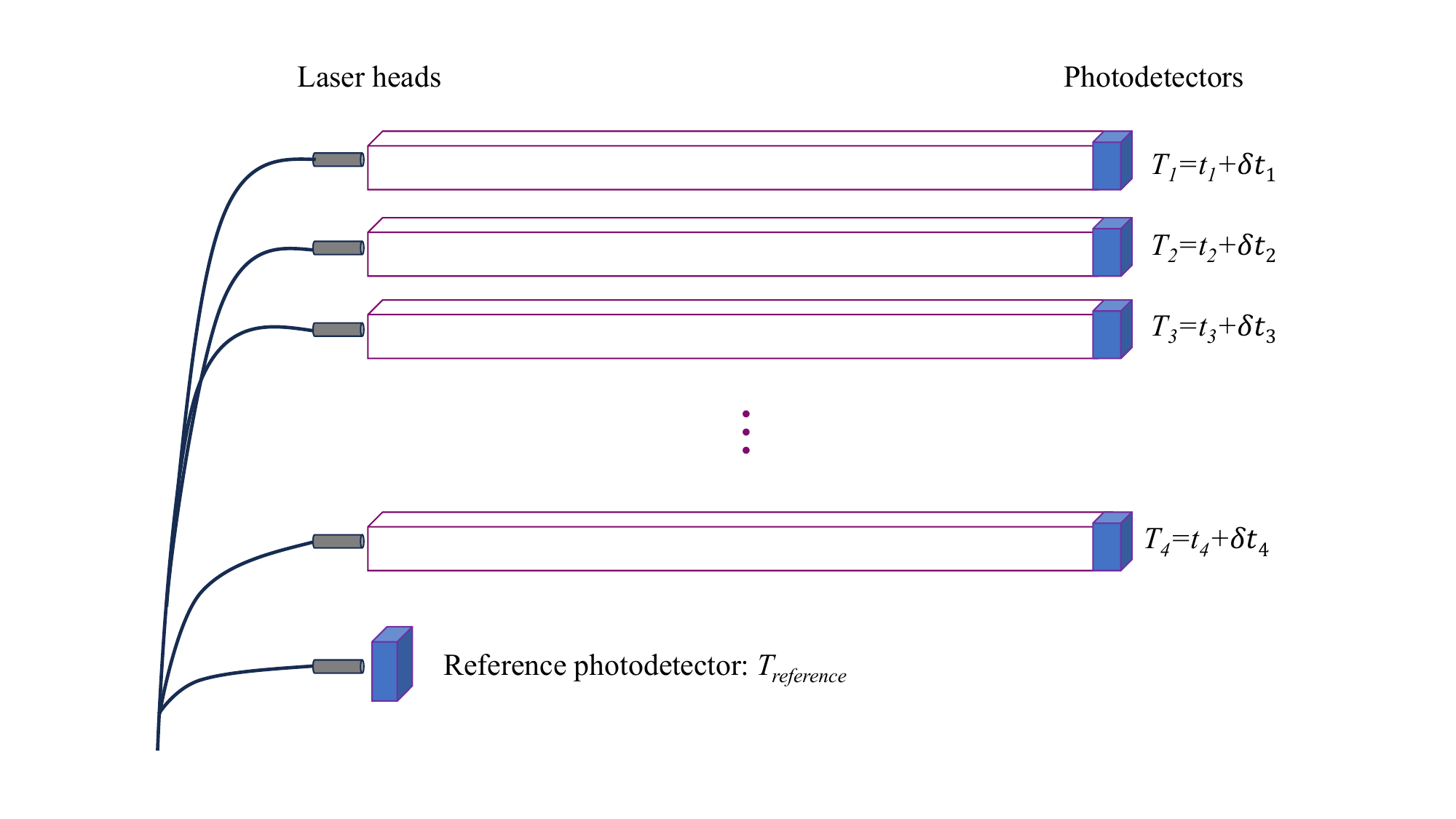} 
\caption{Schematic of the laser calibration system, depicting laser heads positioned on one side and photodetectors on the opposite side, with $T_i$ (where $i = 1,~2,~3, ...,~N$) representing the time recorded by DAQ.}
\label{cali_principle}
\end{figure}

The fundamental concept of the calibration system is illustrated in Fig.~\ref{cali_principle}, which depicts a laser source and $N$ channels of scintillator detectors. Light from the laser splits identically to multiple heads. Each scintillator strip mounts a photodetector on one side and a laser head on another side. According to the R\&D for new scintillators~\cite{HTR}, the numbers of photoelectrons (nPE) obtained in cosmic ray tests are 100-300 for the scintillator detectors. As will be described in Sec.~\ref{section3}, nPE obtained from each laser head is about 150.  For the $i$-th scintillator detector channel, the time recorded by the data acquisition system (DAQ) is given by:
\beq
T_i = t_i + \delta t_i~(i=1,~2,~3,...,N),
\eeq
where time $t_i$ corresponds to the average time of the laser photons arriving the photodetector. To determine $t_i$, one of the laser heads is connected to a reference photodetector. We assume $t_i$ are identical to $T_{\rm ref}$, the time obtained from the reference photodetector. Deviations denoted by $\delta t_i$ arise mainly due to the influence of electronic components. By comparing $T_i$ with $T_{\rm ref}$, these deviations can be quantified, enabling the calibration of the detector.

\begin{figure}[htbp]
\centering
\includegraphics[width=0.8\textwidth]{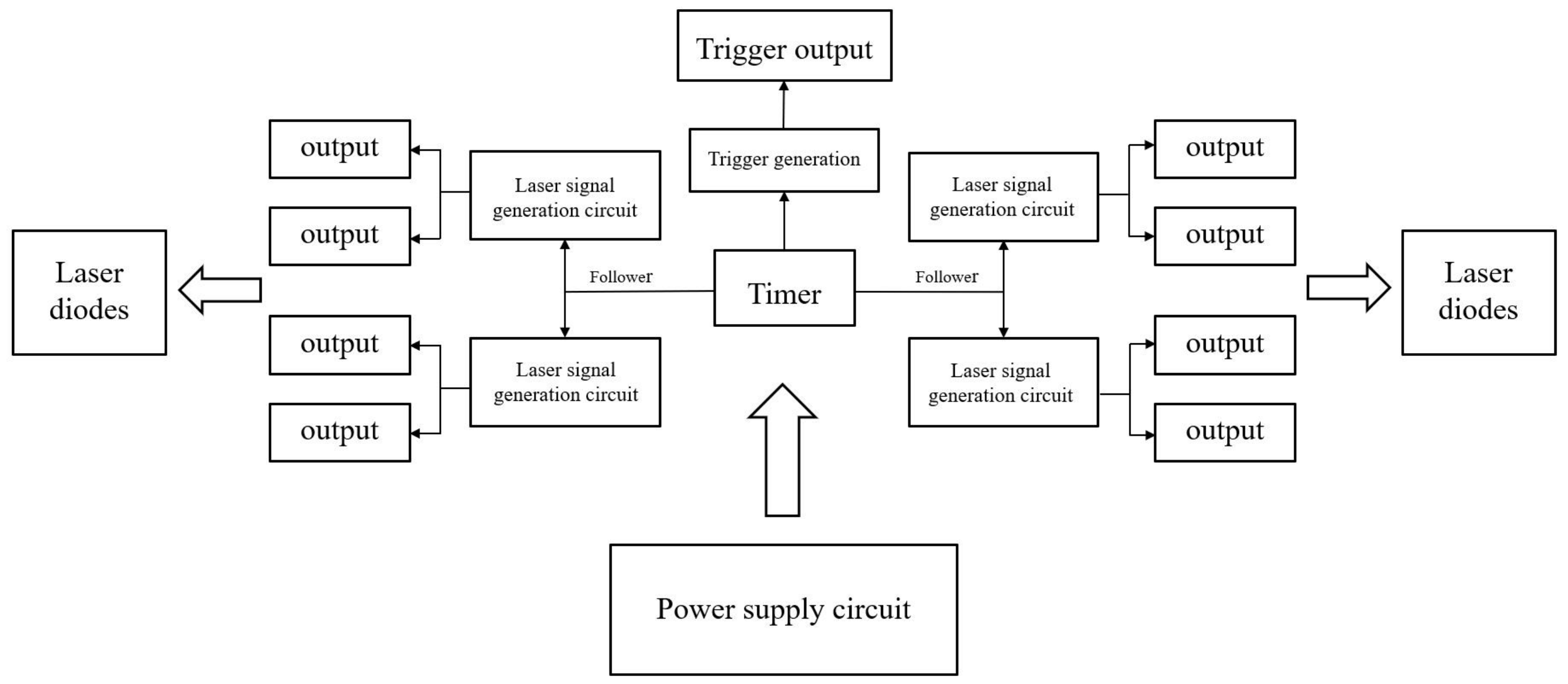} 
\caption{The schematic of the laser driver.}
\label{sg_schematic}
\end{figure}

Figure~\ref{sg_schematic} illustrates the schematic of the laser driver, which generates turn-on signals for the light source. The timer, as the core component, delivers the start signal to four "Laser signal generation circuits". Eventually these four circuits will produce short pulses with a common width of several nanoseconds for each channel. Additionally, the time is connected to a separate chip to generate square waves for the DAQ system, which is the "Trigger output". To ensure excellent time resolution, GaN FETs are used for high-speed switching in the integrated circuits (ICs), enabling the laser to emit short-pulse light. The power supply circuit incorporates a low-dropout regulator (LDO) for the driver circuit and a boost converter for the laser diode.

The laser diodes, chosen as the light source, are mounted on a separate printed circuit board (PCB). To efficiently distribute the high-power laser output, an optical splitter is utilized, directing the light to multiple laser heads for uniform illumination across the system.

\subsection{Selection of laser diode}

Laser diodes are highly efficient, compact, and reliable light sources, widely employed across diverse applications. They are particularly effective in pulse mode and can be operated at low voltage, making them ideal for calibrating optical detectors. In pulse mode, the output power decreases with reduced pulse widths~\cite{lddriver2019}. Consequently, calibration can be performed at low power levels.

We utilized the Osram PLPT5 447KA laser diode~\cite{lddata}, which operates at a wavelength of $445~\nm$ and delivers a maximum output power of $2~\W$. The laser diode exhibits distinct emission angles of $0.17~\rad$ parallel to the PN junction and $0.83~\rad$ perpendicular to it. The laser diode's emission wavelength aligns precisely with the peak PDE of the Hamamatsu S14160 SiPM used in our experimental setup, which will be described in Sec.~\ref{section3}.

\subsection{Design of laser driver}

\begin{figure}[htbp]
\centering
\includegraphics[width=0.8\textwidth]{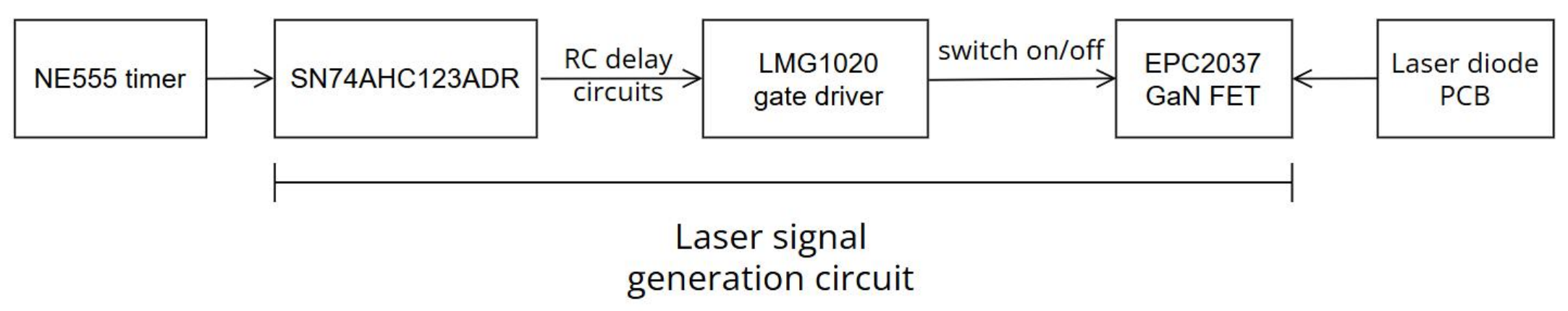} 
\caption{Workflow of single channel. Only critical components are demonstrated.}
\label{workflow}
\end{figure}

Figure~\ref{workflow} illustrates the workflow of the drive circuit for a single channel. An NE555 timer generates square waves at $1~\khz$ and $10~\khz$, which are input into a Texas Instruments SN74AHC123ADR monostable multivibrator~\cite{sn74}. The multivibrator produces dual outputs with pulse widths of hundreds of nanoseconds, which can be adjusted through an external resistor and capacitor. The signals then pass through two RC integrator circuits to the IN+ and IN- ports of the LMG1020 low-side gate driver~\cite{lmg1020}, which features a minimum input pulse width of $1~\ns$ and a maximum output pulse current of $7~\A$. The time constant of the RC integrator circuits adjusts the output pulse width. Two resistors, each with a minimum value of $2~\ohm$, are placed at the gate driver output to limit excessive current flow.

The output of the LMG1020, a short-width pulse, is fed into a switching component. To achieve high-speed switching, GaN FETs are commonly used in laser systems due to their fast turn-on speeds, compact form factors, and low on-resistance, making them highly suitable for drive circuit design~\cite{narrowpulse, highpowernano}. We utilized the EPC2037 GaN FET from Efficient Power Conversion Corporation~\cite{epc2037}, which supports a maximum drain pulsed current of $2.4~\A$.

\begin{figure}[htbp]
\centering
\includegraphics[width=0.45\textwidth]{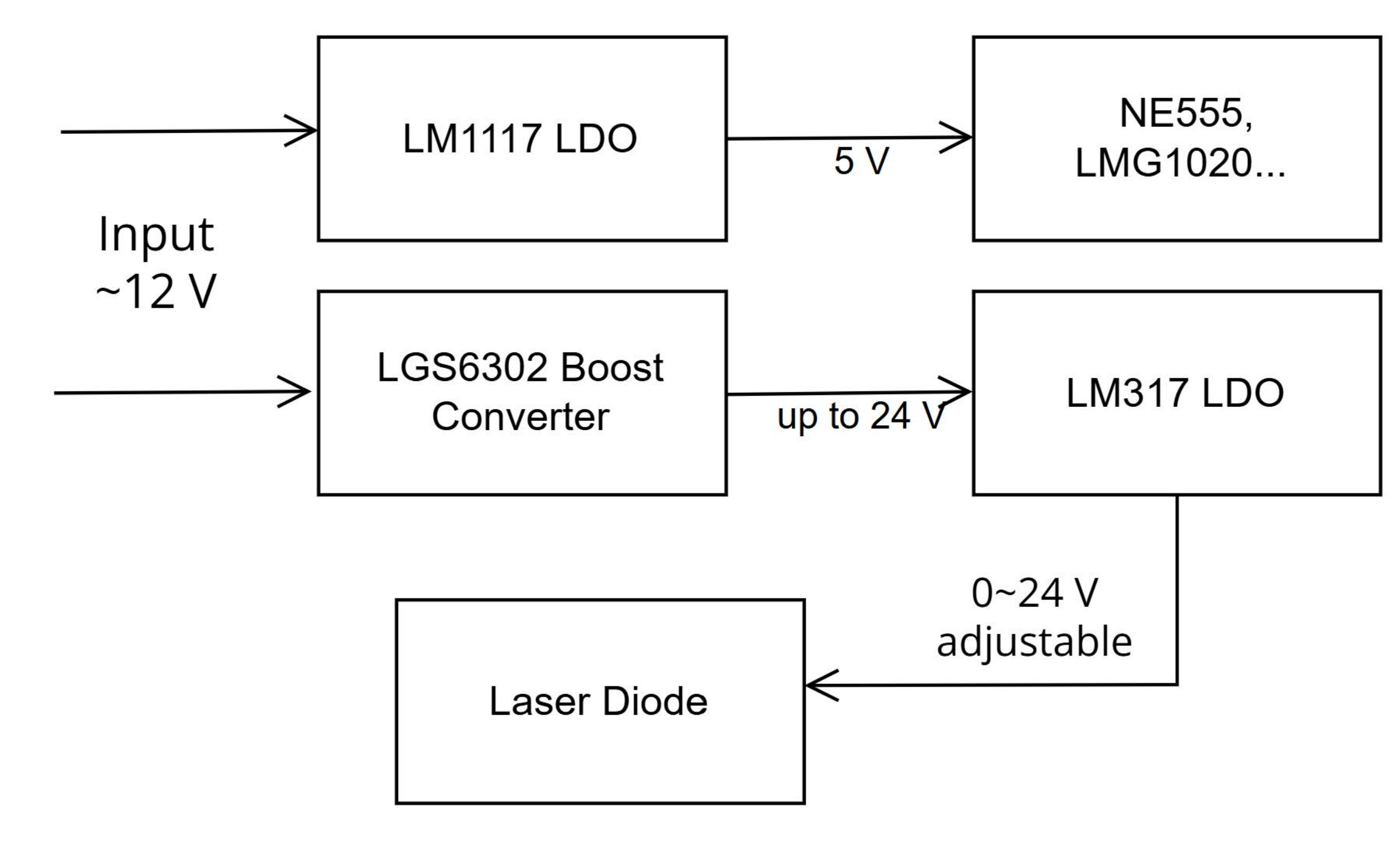} 
\caption{Schematic of the power supply. The system utilizes two sets of power.}
\label{power}
\end{figure}

Figure~\ref{power} illustrates the power supply system, which begins with an input voltage range of $12-15~\V$. The LM1117-5.0 LDO provides a stable $5~\V$ supply for the drive circuit. The LGS6302B5 boost converter steps up the voltage to a maximum of $24~\V$, which is then fed into the LM317 adjustable LDO. The LM317 incorporates a potentiometer, enabling adjustable output voltage to meet the varying operational requirements of the laser diode.

\subsection{Laser diode circuit}

\begin{figure}[htbp]
\centering
\includegraphics[width=0.45\textwidth]{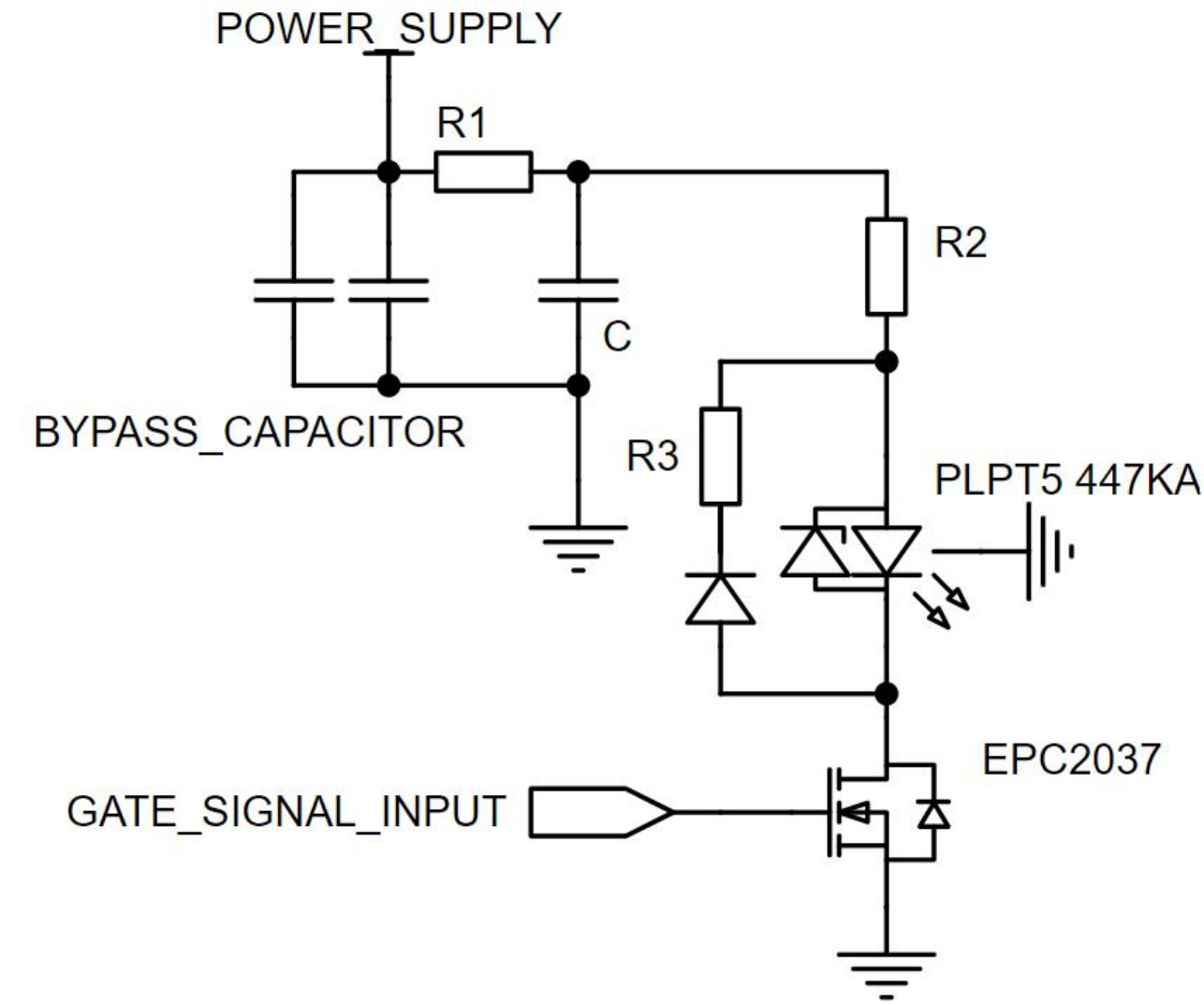} 
\caption{Schematic of the laser diode circuit. The diode and resistor $R3$ are included to minimize the influence of reverse current.}
\label{Schematic_laser}
\end{figure}

Figure~\ref{Schematic_laser} illustrates the laser diode circuit. The capacitor \textbf{C} discharges only upon receiving signals from the gate driver. The discharge rate is determined by the resistor $\textbf{R1} = 300~\ohm$ and capacitor $\textbf{C} = 500~\pF$, enabling a laser output repetition frequency of $1.33~\mhz$~\cite{highside}.

Upon receiving the trigger signal, \textbf{R2}, \textbf{C}, and the laser diode collectively form a discharging circuit that operates as a conventional series \textbf{RLC} circuit. The \textbf{RLC} discharging circuit should work in an underdamped state. In such a state, when there is an initial voltage, the discharge current of the capacitor can be expressed as
\beq\label{eq_1}
i(t) = I_0 e^{-\alpha t} \sin{\omega_d t}
\eeq
with $\alpha=R/2L$, $\omega_0 = 1/LC$, and $\omega_d = \sqrt{\omega^2_0 - \alpha^2}$. Here, $C$, $L$, and $R$ are the discharging capacitor, the total inductance, and the total resistance, respectively, while $\omega_0$ influences pulse width. Applying this formula to our design, the $C$ should be the discharging capacitor \textbf{C}, the $R$ should be the sum of \textbf{R2} and equivalent resistance, and the $L$ is equivalent inductance.

To achieve a narrow pulse width with sufficient current, careful optimization of the circuit configuration and PCB layout is crucial. Generating a short pulse requires a sufficiently large $\omega_d$, which can be achieved by minimizing the inductance $L$. Additionally, a small $L$ helps to suppress current oscillations, ensuring the emission of a single, well-defined short light pulse during the discharge process. To minimize inductance, wires should be kept as short as possible. Furthermore, a diode and a resistor are connected in parallel across the laser diode to form a clamping circuit~\cite{highside}, as shown in Fig.~\ref{Schematic_laser}. This clamping circuit limits reverse current and absorbs harmful voltage spikes, protecting the laser diode from damage caused by oscillations.

\section{Prototype development and performance analysis}
\label{section3}

\begin{figure}[htbp]
\subfigure[]{
\begin{minipage}[]{\linewidth}
\centering
\includegraphics[width=0.5\textwidth]{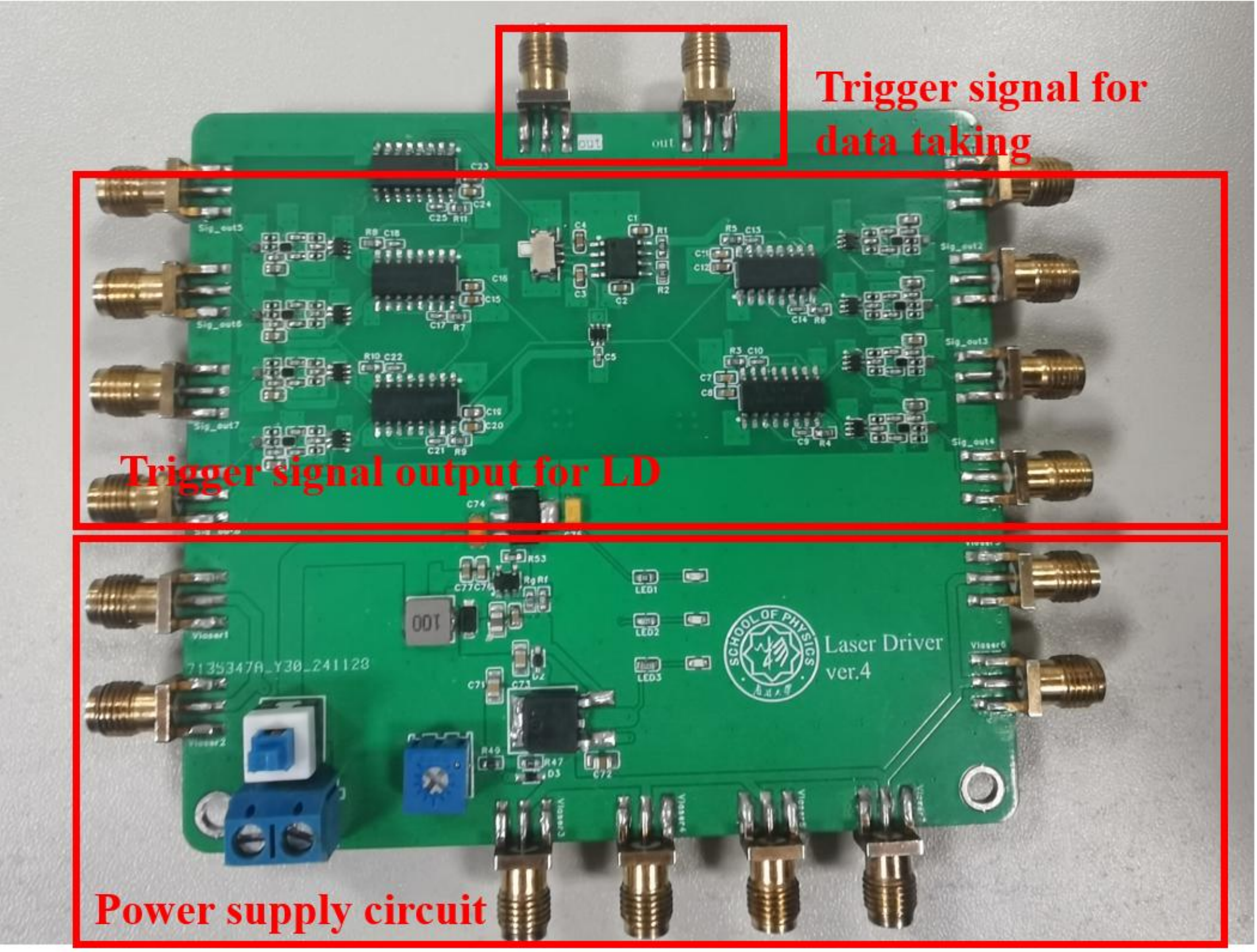} 
\label{prototype}
\end{minipage}}
\subfigure[]{
\begin{minipage}[]{\linewidth}
\centering
\includegraphics[width=0.3\textwidth]{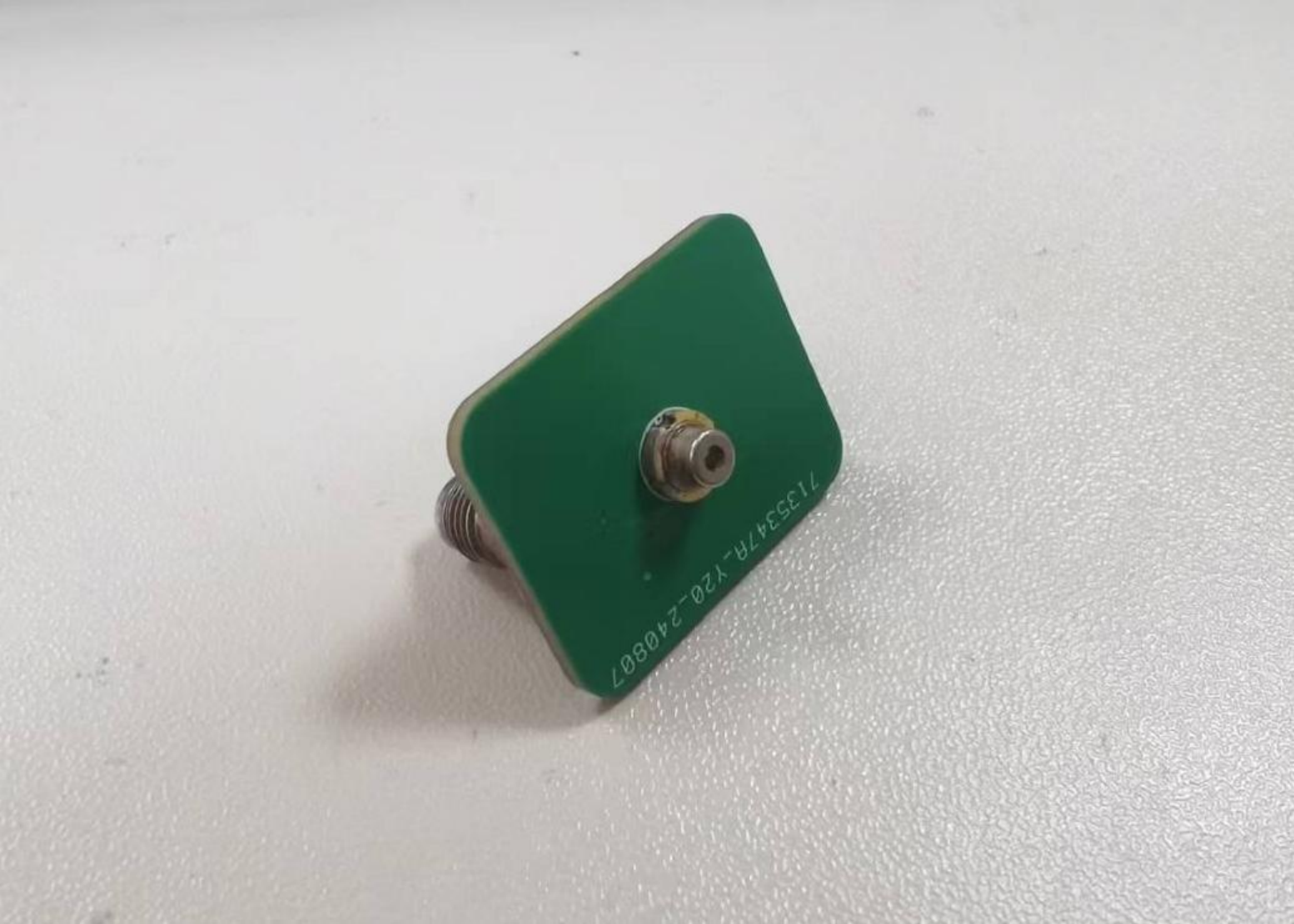}
\includegraphics[width=0.3\textwidth]{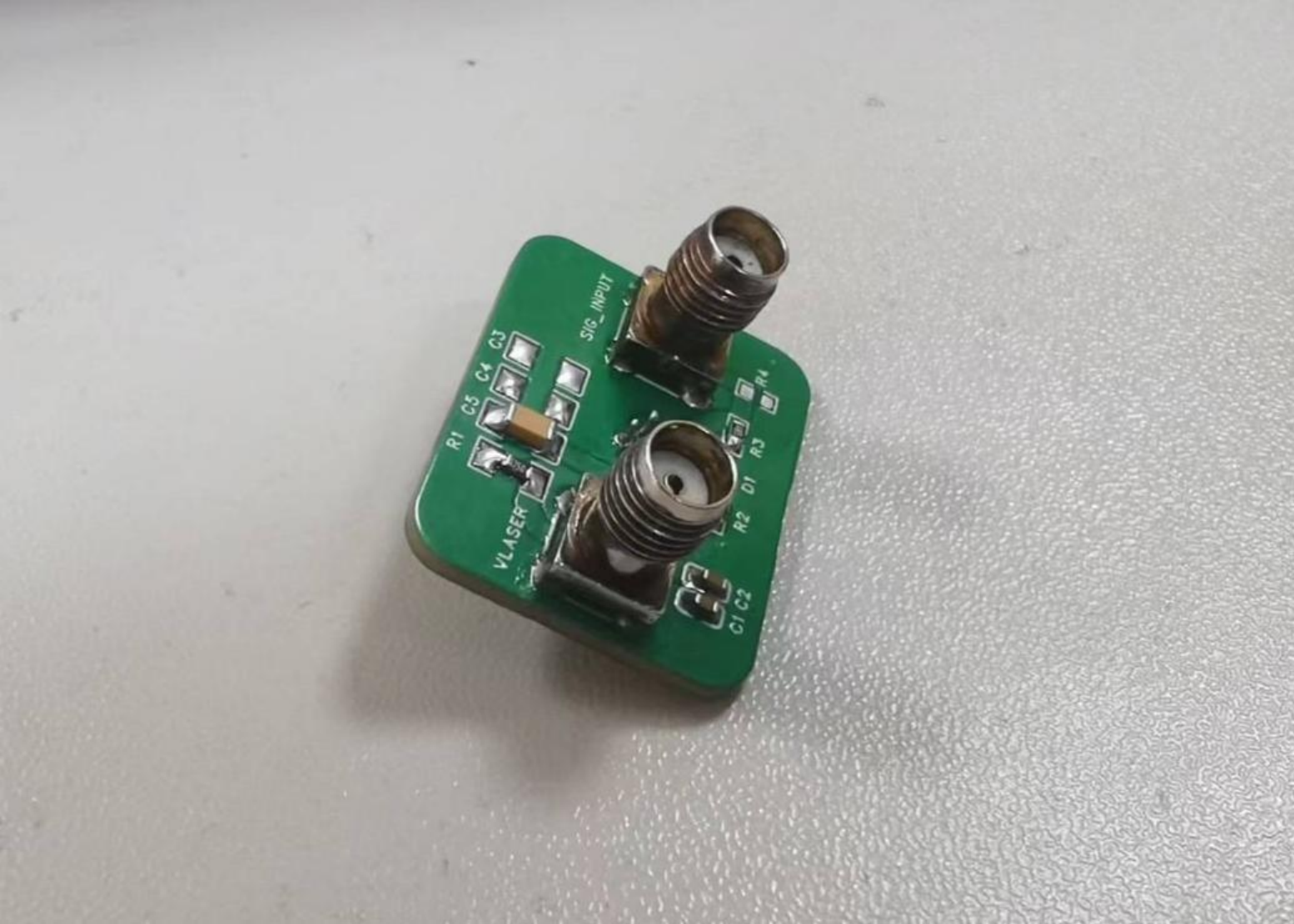}
\end{minipage}
\label{laserpcb}}
\caption{(a) PCB of the laser driver, from top to bottom are the trigger output for external use, the laser trigger signal for the laser diode, and the power supply for the laser diode. (b) Laser diode mounted on a small PCB. }
\label{PCBs}
\end{figure}

We developed a prototype of the laser system and used the scintillators developed for the KLM upgrade~\cite{HTR}, as illustrated in Fig.~\ref{PCBs}. The scintillator bars have the same geometry $2~\cm \times 4~\cm \times 100~\cm$, while the one reported in Ref.~\cite{HTR} has a geometry of $2~\cm \times 4~\cm \times 135~\cm$. The system comprises two main components: the laser driver PCB, measuring $97~\mm \times 86~\mm$, and the laser diode mounted on a smaller PCB measuring $26~\mm \times 21~\mm$. The laser driver is designed with eight channels for laser control, two channels for trigger output, and an additional eight channels dedicated to power delivery to the laser diodes.

To evaluate the system's performance, two critical aspects must be addressed. First, to determine the time deviation of a scintillator-based detection channel accurately, the time resolution of the corresponding calibration system channel, such as illustrated in Fig.~\ref{cali_principle}, must surpass that of the detector channel. For the KLM upgrade, R\&D demonstrates that the time resolution of a detector channel can be better than $100~\ps$, and in some cases, around $50~\ps$~\cite{HTR}. Therefore, the calibration process must achieve the highest possible time resolution for a single channel. Second, the intrinsic deviations among the multiple channels of the laser driver must be minimized. These deviations, arising from variations in the performance of calibration channels, can significantly compromise the accuracy of the system calibration. Consequently, precise self-calibration of the laser driver is essential to mitigate these deviations.

Considering these two aspects, several tests have been conducted. We selected the Hamamatsu S14160 series to construct a $1\times 4$ SiPM array for photon detection. These SiPMs are connected in parallel on a custom-designed PCB. A preamplifier specifically designed for SiPM signals~\cite{PreAMP} provides a gain of +26 dB and a bandwidth of $426~\mhz$, enabling a $\Trise$ of $1~\ns$ while reducing the baseline noise level to as low as $0.6~\mV$, which is significantly lower than the typical pulse height over $500~\mV$ of a calibration signal. For data acquisition, we employed the DT5742 digitizer~\cite{DT5742}, which features 16 readout channels and an external trigger channel. Each readout channel has a bandwidth of $500~\mhz$ and supports analog signal sampling rates of up to $5~\ghz$. The data were processed using the Constant Fraction Discriminator timing method, implemented by fitting to the leading edge of the signal and extracting the timing with a fraction of $f = 0.2$.

\subsection{Performance evaluation}

\begin{figure}[htbp]
\centering
\includegraphics[width=0.45\textwidth]{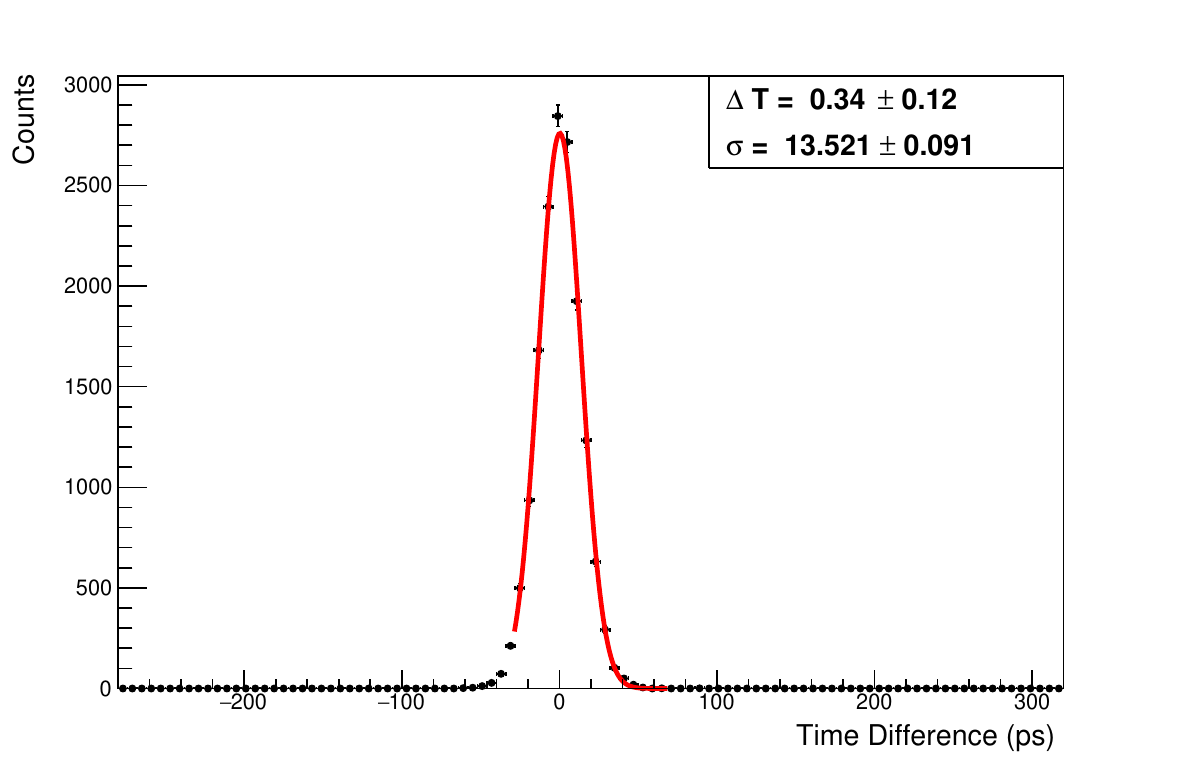}
\caption{Time difference between two SiPM arrays in the test on the laser driver circuit and the laser diode, which has the time resolution for $13.52 \pm 0.09~\ps$.}
\label{t12}
\end{figure}

In the first step, we tested the laser driver circuit and the laser diode. The experimental setup consists of a single laser diode and two SiPM arrays, with the laser diode positioned directly facing the arrays. Neutral density filters were employed to regulate the laser's output power and prevent detector saturation. The result is shown in Fig.~\ref{t12}. The time resolution of the difference is $13.52 \pm 0.09~\ps$. It reflects the performance of the LD and its driver circuit. Such resolution is only achievable with a laser diode capable of emitting sufficient short and intense pulses. This result confirms the system's effectiveness as a precise calibration source.

\begin{figure}
\centering
\includegraphics[width=0.6\textwidth]{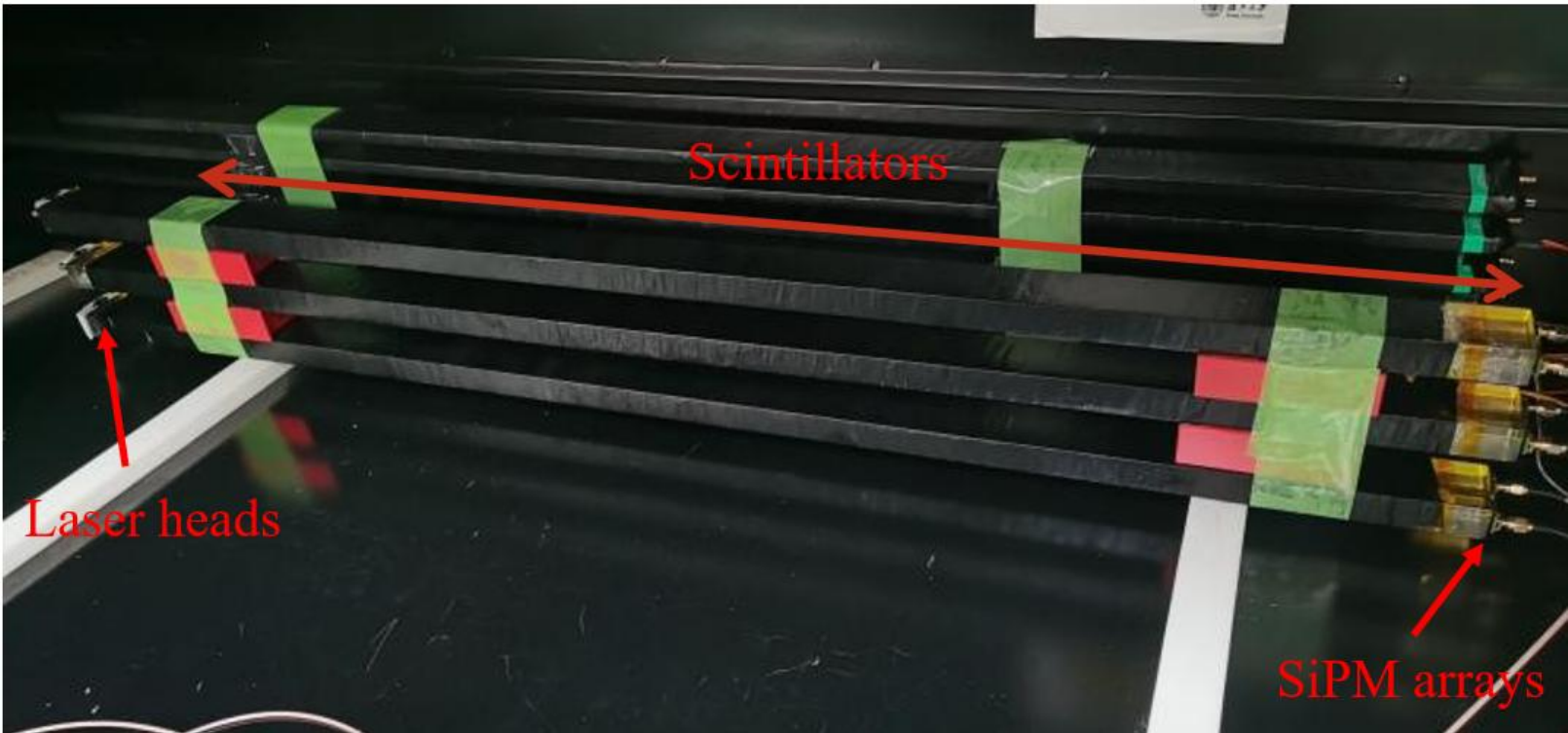} 
\caption{Experimental setup featuring three scintillator bars. Fiber heads are mounted on the left side for light input, while SiPM arrays are positioned on the right side for signal detection.}
\label{setup}
\end{figure}

\begin{figure}[htbp]
\centering
\includegraphics[width=0.4\linewidth]{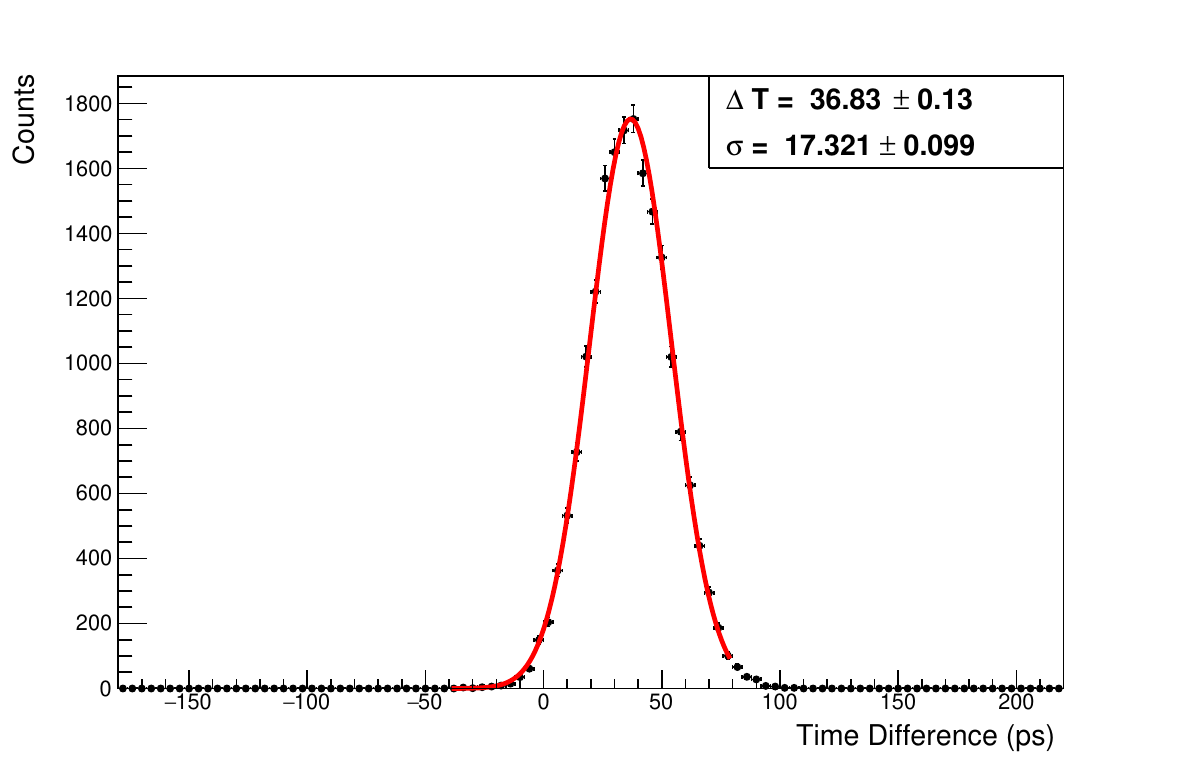}
\includegraphics[width=0.4\linewidth]{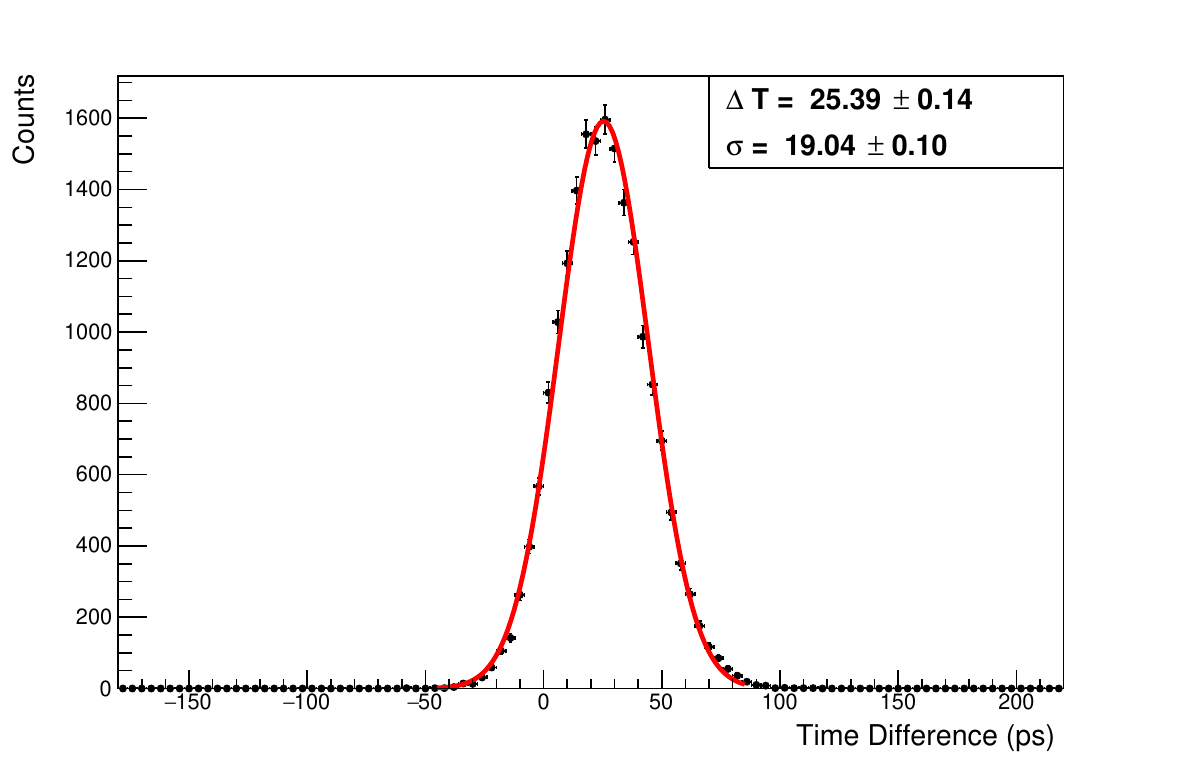}
\caption{(a) Time difference between $T_1$ and $T_2$. (b) Time difference between $T_3$ and $T_2$.}
\label{result1}
\end{figure}

To further evaluate the laser system's performance, we test it with scintillator detectors, as shown in Fig.~\ref{setup}. In this setup, neutral density filters are replaced with a $1\times 8$ optical splitter, mounted to the laser diode via a flange. The flange directs light into a 1-meter multi-mode fiber (working wavelength: $850~\nm$, core diameter: $62.5~\um$) with an \textbf{FC} connector, positioning close to the laser diode for optimal coupling efficiency. The splitter divides the light signal, transmitting it through the fiber to high-transparency scintillators from GaoNengKeDi Company, known for their exceptional timing performance in the R\&D for KLM upgrade~\cite{HTR}. The setup includes three scintillation bars, each wrapped in aluminum foil and black tape to maximize light collection efficiency. One fiber head is attached to one end of each scintillator bar for optimal light transmission using a 3D-printed compartment, while a SiPM array is positioned at the opposite end. The laser diode operates at a nominal voltage of $13~\V$, adjustable for higher power output as needed. The results are presented in Fig.~\ref{result1}. With $T_2$ as the reference offset, the calibration constants for channels 1 and 3 are $-37~\ps$ and $25~\ps$, with resolutions of $17~\ps$ and $19~\ps$, respectively. This yields a time resolution of about $13~\ps$ for a single calibration channel.

\subsection{Self-calibration of the drive circuit channels}

\begin{figure}[htbp]
\centering
\includegraphics[width=0.4\textwidth]{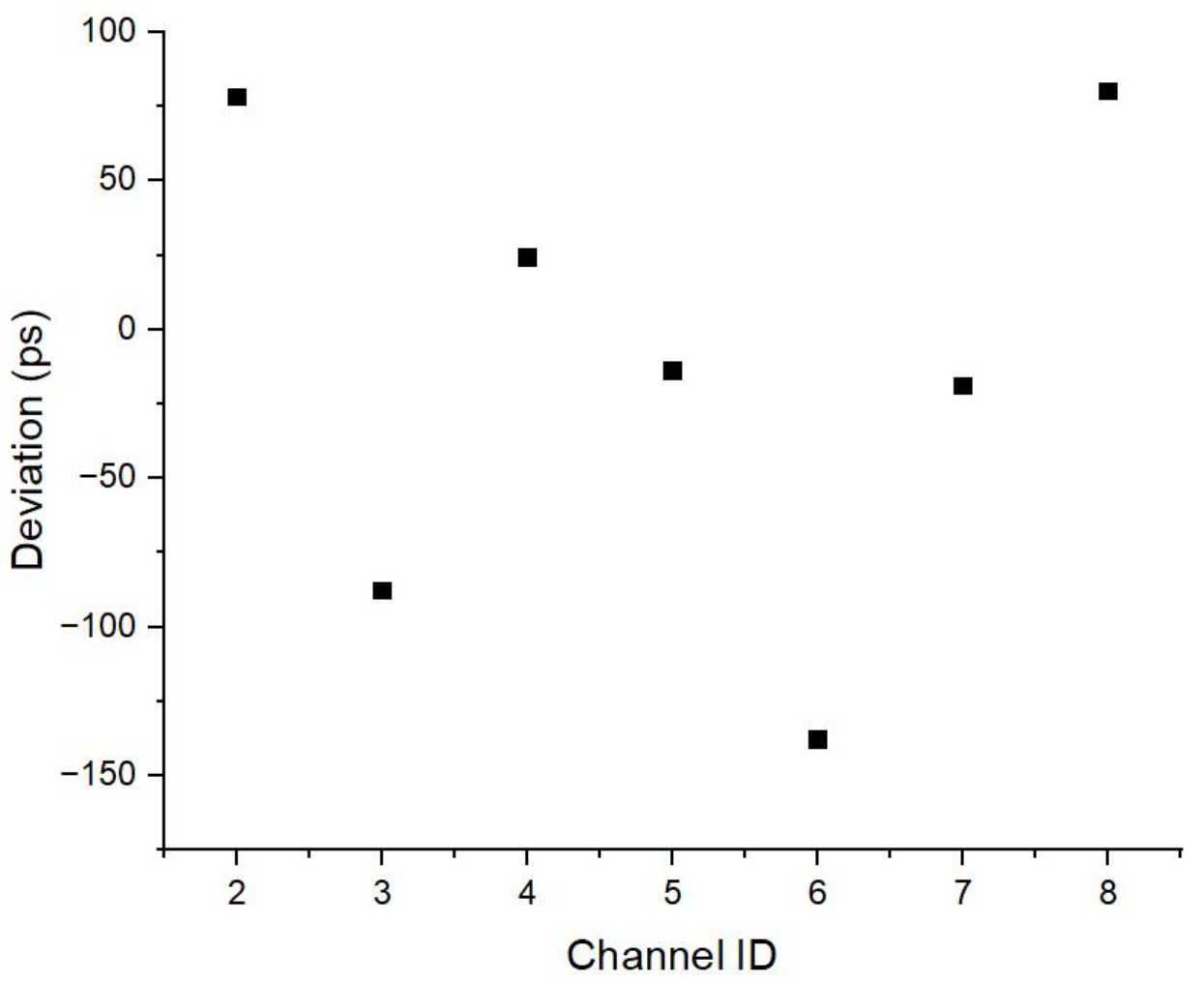}
\caption{Inter-channel deviations compared to the first calibration channel in the laser driver.}
\label{deviation}
\end{figure}

We analyze deviations among the eight laser control channels in the laser driver. In the test, the laser driver remains fully functional, with the sole modification involving the interchange of control channels connected to the same laser diode. The SiPM array employed is identical to that used in prior tests; however, only a single SiPM array is utilized in this configuration. Signals from the SiPM array are compared with square waves from "Trigger output" depicted in Fig.~\ref{sg_schematic} to determine the absolute deviations. The first laser control channel served as the reference for calculating relative deviations. As shown in Fig.~\ref{deviation}, the remaining seven channels are compared against the first, revealing a maximum deviation of $138~\ps$, with most values within $100~\ps$. The deviations among all the calibration channels are less than $250~\ps$.

\section{Conclusion}

In summary, we have successfully developed a prototype laser calibration system tailored for the Belle II experiment upgrade, addressing the stringent requirement for time calibration in its large-size KLM Detector. By leveraging a laser diode as the light source and integrating a fast pulse laser drive circuit with high-speed switching GaN FETs and gate drivers, the system achieves an exceptional timing resolution of approximately $13~\ps$ on the single channel. The deviations among the calibration channels are less than $250~\ps$. These results, obtained through comprehensive evaluations with scintillators, demonstrate the potential to meet the demanding precision needed for tens of thousands of scintillator channels.

\section*{Acknowledgment}

This work is partially supported by the National Key R\&D Program of China under Contract No. 2022YFA1601903; the National Natural Science Foundation of China under Contracts Nos. 12175041, 12405099.

\end{document}